\documentclass[aps,pre,showpacs,superscriptaddress,twocolumn]{revtex4}
\usepackage{graphicx,psfrag}

\begin{document}
\title{Formation of Liesegang patterns in the presence of an electric field}

\author{I. Bena}
\affiliation{Department of Physics, 
University of Gen\`eve, CH-1211 Gen\`eve 4, Switzerland}
\author{M. Droz}
\affiliation{Department of Physics,
University of Gen\`eve, CH-1211 Gen\`eve 4, Switzerland}
\author{Z. R\'acz}
\affiliation{Institute for Theoretical Physics - 
HAS, E\"otv\"os University, P\'azm\'any s\'et\'any 1/a, 1117 Budapest, Hungary}

\date{\today}

\begin{abstract}
The effects of an external electric field on the formation of 
Liesegang patterns are investigated. The patterns are assumed 
to emerge from
a phase separation process in the wake of a diffusive reaction front. 
The dynamics is described by a Cahn-Hilliard equation with a moving 
source term representing the reaction zone, and the electric field 
enters through its effects on the properties of the reaction zone. 
We employ our previous results
[I. Bena, F. Coppex, M. Droz, and Z. R\'acz, J. Chem. Phys. {\bf 122},
024512 (2005)] on how the electric field changes 
both the motion of the front, as well as the amount of reaction 
product left behind the front, and our main conclusion is that 
the number of precipitation bands becomes finite in a finite electric field. 
The reason for the finiteness in case when the electric field drives 
the reagents towards the reaction zone is that the width of consecutive 
bands increases so that, beyond a distance $\ell_+$, 
the precipitation is continuous (plug is formed). 
In case of an electric field of opposite polarity, 
the bands emerge in a finite interval $\ell_-$, 
since the reaction product decreases with time and the conditions for
phase separation cease to exist. 
We give estimates of $\ell_{\pm}$ in terms of 
measurable quantities and thus present an experimentally 
verifiable prediction of the "Cahn-Hilliard equation with a  moving source" 
description of Liesegang phenomena.

\end{abstract}
\maketitle

\section{Introduction}
\label{introduction}

Precipitation patterns named after Liesegang~\cite{liesegang1896,henisch91}
have been investigated for more than 100 years.
Their discovery came from work on photoemulsions~\cite{liesegang1896},
and the interest in these patterns was sustained during all these years
by recognizing that the underlying dynamics had connections to
rather diverse physical- (e.g. near-equilibrium crystal growth~\cite{krug}),
geological- (formation of agates~\cite{kruhl}), chemical- (pattern formation
in reaction-diffusion systems~\cite{swinney}) and more exotic (e.g. aggregation of
asphaltene in crude oil \cite{Sheu2004}) phenomena.
From theoretical point of view, the Liesegang patterns
enjoyed continuing attention since attempts at their explanations were
testing the theories of 
precipitation processes~\cite{Ostwald,Wagner1950,Prager1956}
(for a recent overview see Ref.\cite{antal98})
and, furthermore, the phenomenon was considered as a
highly-nontrivial example of pattern formation in the wake of a
moving front~\cite{Dee1986,cross94}.

The origin of recent attention to Liesegang patterns is
the hope that the
phenomena may be relevant in engineering of meso- and microscopic
patterns~\cite{Giraldo2000,Lebedeva2004,Fialkowski2005}. The novelty
of the idea is
that, in contrast to the "top-down" processing (removing material in
order to create a structure),
the Liesegang dynamics provides a "bottom-up" mechanism where the
structure emerges from a bulk precipitation process~\cite{Lebedeva2004}.
Of course,
many obstacles will have to be overcome before such a strategy
succeeds, the main one being the problem of controlling the pattern
generated by a reaction-diffusion process. It is known experimentally,
with the results formulated in the Matalon-Packter law \cite{Matalon1955},
that some degree of control may be exercised through the appropriate
choice of the concentrations
of the inner and outer electrolytes participating
in the process. It is also known, but much less understood, that the
gel strongly influences the resulting patterns \cite{Toramaru2003}.
Furthermore, recent experiments \cite{Fialkowski2005}
indicate that the shape of the gel may be used in designing appropriate
geometry in the precipitation patterns.

\begin{figure}[h!]
\begin{center}
\vspace{1.5cm}
\includegraphics[width=0.9\columnwidth]{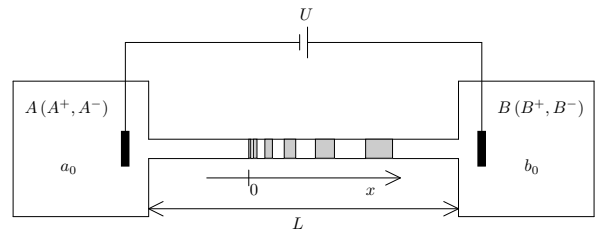}
\end{center}
\caption{Schematic representation of the system under study. 
The inner electrolyte $B(B^+,B^-)$ is dissolved in a gel matrix 
inside a part of the small central cylinder 
(from point ``$0$" to the left on the $x$-axis).
The outer electrolyte $A(A^+,\,A^-)$, of much higher concentration, 
is brought in contact with the gel 
(at point ``$0$" on the $x$-axis).
The two reservoirs of electrolytes $A(A^+,\,A^-)$ and $B(B^+,B^-)$ 
assure constant 
concentrations $a_0$ and $b_0$ ($a_0\gg b_0$)
of the ions at the two ends of the central cylinder.
The outer electrolyte ions diffuse into the gel, 
where the reaction
$A^-+B^+\rightarrow C$ takes place.
(The background ions $A^+$ and $B^-$ do not react.) 
The reaction front moves {\em without convection} 
towards the right, and precipitation bands -- i.e.,
the alternation of high-density-$C$ regions (shaded areas)
and low-density-$C$ regions -- emerge inside 
the cylinder in the wake of this moving reaction front.
A tension $U=V_B-V_A$ is applied 
between the ends of the central cylinder of length $L$,
which corresponds to an average applied field intensity $E=-U/L$.
A positive tension $U>0$ 
(corresponding to $E<0$)
drives the reacting ions towards the reaction zone
({\em forward} electrc field).
A  negative tension  $U<0$ 
(corresponding to $E>0$) works against the reacting ions reaching the
reaction zone  ({\em reverse} electric field).}
\label{figure1}
\vspace{1.5cm}
\end{figure}

The methods of control described above
are somewhat rigid since the parameters
cannot be changed during the process while, ideally, one requires
an easily tuned, flexible external field for control.
In principle, the electric field provides such an external control and,
indeed, there is experimental 
evidence~\cite{happel29,kisch29,ortoleva78,sharbaugh89,das90,das91,das92,sultan00,sultan02,ghoul03,lagzi02,lagzi03,shreif04}
that an electric field significantly alters the emerging pattern
(see Fig.\ref{figure1} for a schematic experimental setup).
Unfortunately, these experiments
give only a qualitative picture about the effects of the
electric field, and the theory is even 
less developed~\cite{ortoleva78,ghoul03,lagzi02,lagzi03,Feeney-Orto1983}.

Our aim with this paper is to improve the theoretical description of
electric field effects and to bring it up to the level of quantitative
predictions. We have shown recently \cite{bena05}
that there are problems with previous 
attempts~\cite{ortoleva78,ghoul03,lagzi02,lagzi03,Feeney-Orto1983} 
which incorporate 
the electric field by
assuming that it results in the drift of the reacting ions as well as
in the drift of the reaction zone. Treating the background ions
properly by using the electroneutrality condition, we found that
the main effect of the field is that the amount of
reaction product left behind the reaction zone increases (decreases)
linearly in space depending whether the field drives the reacting
ions towards (away) from the reaction zone. Building on these
results, we shall show below using the "Cahn-Hilliard 
equation with moving source" model \cite{antal99,antal01}, 
how the field affects
the precipitation pattern itself.

The choice of the model must be explained since a century of research
did not lead to a generally accepted theory of this pattern forming
process. The reason is perhaps the complexity of
the interplay between
the motion of the reaction front and the precipitation dynamics of the 
reaction product (and the intermediate reaction steps that
may also be present), thus preventing the
creation of a single model encompassing all the possibilities. The
approach we consider \cite{antal99,antal01}
simplifies the situation by
first treating the kinetics of reaction and the motion of the
reaction front \cite{galfi88,cornell93,unger00}. Then,
the reaction product generated by the front
is inserted as a source in the phase separation process described by the
Cahn-Hilliard equation~\cite{gunton83,cahn58,cahn61,hohenberg77}.
This theory has been shown \cite{antal99,antal01,racz99,droz00,magnin00}
to generate Liesegang patterns which satisfy
the time- and spacing-laws \cite{morse,jabli} (patterns
emerging from most other theories do the same),
the Matalon-Packter law \cite{Matalon1955} (significantly
fewer theories can produce such patterns),
and the width law~\cite{Muller82,droz99,racz99}
(no other theory yields this law since none of them has an underlying
thermodynamics for setting the values of steady-state concentrations).
This theory is also distinct from previous approaches in that
it has a minimum number of parameters which can be related to
experimentally measurable quantities and thus the theory can give
quantitative predictions \cite{racz99}. We should also note that,
depending on the motion of the front, the phase separation process
may occur at the position of the front or well behind the front
\cite{antal01} thus the model can also describe the limit of small imposed
gradients \cite{Muller-Ross03}.

In order to summarize the results obtained from the above theory for
the case of external field present, let us first describe the setup
(Fig.\ref{figure1}) more precisely.
A chemical reagent $B\equiv (B^+,B^-)$, called inner electrolyte, 
is dissolved in a gel matrix
inside a part of the small central cylinder (from point ``$0$" to the
right on the $x$-axis).
A second reactant $A\equiv (A^+,A^-)$ (outer electrolyte), 
of much higher concentration, is brought in contact with the gel
(at point ``$0$" on the $x$-axis).
The two reservoirs of electrolytes $A$ and $B$ are providing
constant concentrations $a_0$ and $b_0$ ($a_0\gg b_0$)
of the reagents at the two ends of the central cylinder.
The outer electrolyte $A$ diffuses into the gel and reacts
($A^-+B^+\rightarrow C$) with the inner electrolyte $B$.
The reaction front moves to the right and,
under appropriate conditions, the reaction product $C$ precipitates,
and one observes the emergence of bands of precipitate
perpendicular to the direction of motion of the
front. The reservoirs of electrolytes $A$ and $B$
are kept at a constant potential difference $U=V_B-V_A$,
which corresponds to an average applied field intensity $E=-U/L$
inside the central cylinder of length $L$.
Let us note that throughout this paper we shall call 
{\em forward electric field} a field 
{\em that drives the reacting ions towards the reaction zone};
in our setup this corresponds to a  positive tension $U>0$, 
or to $E<0$. On the contrary, a field that 
{\em works against the reacting ions reaching the
reaction zone} will be referred to as a {\em reverse electric field},
which in our setup corresponds to
a  negative tension  $U<0$, 
or to $E>0$.

Without the external field ($U=0$), the positions of the
bands ($x_n$, measured from the initial contact of the reagents)
form a geometric series, $x_n\sim (1+p)^n$, where $1+p$ or $p>0$
is called the spacing coefficient. For small fields ($|E|=|U|/L\lesssim 2$ V/m,
for which one has a `sufficient' number of bands, i.e., $\geq 20$),
the band spacing can still be described as geometric series and our first
result pertains to the dependence of the effective spacing
coefficient on the applied field, $p=p(U/L)$. 
We call it effective spacing coefficient
because these geometric series are finite as evidenced by 
the results for higher fields. The reason for the finite number of bands
for electric field that drives the reagents towards the reaction zone
(called forward field in the following) is that 
the width of consecutive bands increases faster 
than the distance between the bands. Thus
one finds that the precipitation is continuous 
(plug is formed) beyond a distance $\ell_+$. 
The number of band is also finite and they
appear in an interval $[0,\ell_-]$ for the case
of a field of reverse polarity. The reason for finiteness,
however, is different.
In case of a field working against the reacting ions 
reaching the reaction zone, the amount of reaction product
generated in the reaction zone
decreases with time and the conditions for phase separation cease
to exist.
The quantities $\ell_+$ and $\ell_-$ are easily accessible in experiments,
and our main result (apart from qualitative observations)
is that we give an estimate of them
in terms of measurable quantities. Thus we provide a way of designing the
spatial range of the emerging pattern. In view of the competing theories
of Liesegang phenomena, this also gives yet another way of
discovering which is the correct description.

Below we present the details in the following order.
First, the Cahn-Hilliard equation and the changes in the source term
in the presence of an electric field are discussed (Sec.~II).
Next (Sec.~III), the results of
numerical simulations and qualitative arguments concerning 
the characteristics of the Liesegang patterns for a forward
applied field are presented.
The case of the reverse polarity is described in Sec.~IV.
The comparison with experiments is discussed in Sec.~V, while
the conclusions
and perspectives are given in Sec.~VI.
Finally, a discussion on the choice of the parameters of our 
theoretical model, as inferred from experimental data, can be found
in the Appendix.

\section{Phase separation and dynamics of the reaction product $C$ 
in the presence of an external electric field}

\begin{figure}[htbp]
\begin{center}
\vspace{1.5cm}
\psfrag{F(m)}{{\Large ${{\cal F}(m)}$}}
\includegraphics[width=0.9\columnwidth]{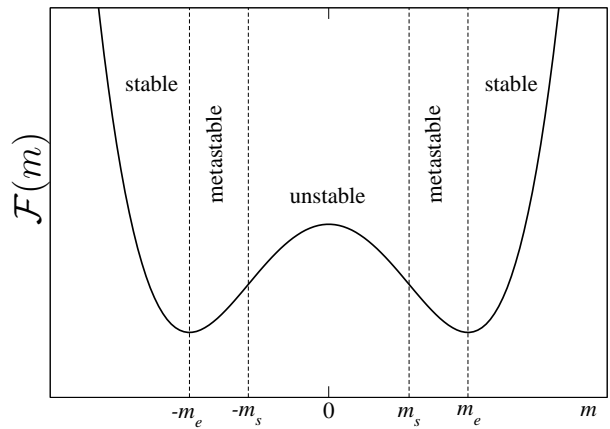}
\end{center}
\caption{The homogeneous part of the free energy density ${\cal F}$ as a function of the reduced density $m$.
The phase separation is an activated process in the metastable regimes $m_s=1/\sqrt{3} < |m| < m_e=1$,
while it goes by spinodal decomposition in the linear instability region $|m| <
m_s=1/\sqrt{3}$.}
\label{figure2}
\vspace{1.5cm}
\end{figure}

There is experimental evidence that the Liesegang pattern is 
an alternation of 
high-density $c_h$  and low-density $c_l$ regions 
of the  precipitate $C$. 
Thus, the reaction product 
phase-separates behind the reaction front, 
and since this takes place on a macroscopic scale, its dynamics can be 
represented through an extension of the Cahn-Hilliard equation 
(in other contexts, this is the equation of model B of critical
dynamics~\cite{hohenberg77}). 
This equation, however, requires the knowledge of the 
free energy density ${\cal F}$
of the system; note that for a homogeneous system, ${\cal F}$ has to 
present two minima in order to accommodate 
the two equilibrium states of high $c_h$
and low $c_l$ densities. The simplest form of ${\cal F}$ having this property 
and containing the minimum number of parameters (which, moreover, guarantees the stability
of the system against short-wavelength fluctuations) is the Ginzburg-Landau 
free energy density
\begin{equation}
{\cal F}[m]=-\frac{1}{2}\varepsilon m^2 +
\frac{1}{4}\gamma m^4 +\frac{1}{2}\sigma (\nabla m)^2\,.
\label{free}
\end{equation}
Here we introduced  the `reduced' concentration
\begin{equation}
m(x,t)=\frac{c(x,t)-(c_h+c_l)/2}{(c_h-c_l)/2}
\end{equation}
that varies between $-1$ and $+1$ when the concentration of the 
product $C$ varies between $c_l$ and $c_h$. 
Note the underlying assumption of the one-dimensional
character of the system, i.e., the fact that all 
the relevant parameters are only $x$-dependent, 
that is well-justified for the experimental 
setup in Fig.~\ref{figure1}.
The parameters $\varepsilon$, $\gamma$, 
and $\sigma$ are system and temperature dependent, 
with (i) $\varepsilon >0$ ensuring that the system is in the 
phase separating regime
(i.e., the temperature is smaller than the critical one); 
(ii) $\gamma = \varepsilon$
in order to ensure that the minima of the free energy correspond to $m=\pm 1$
(the homogeneous high and low density phases of $C$); (iii) $\sigma >0$ in order to 
provide stability against short-wavelength inhomogeneities. 
Figure~\ref{figure2} offers a schematic representation of the 
homogeneous part of the free
energy ${\cal F}$ as a function of $m$, with the different stability regions.
Because the $C$-s are assumed to be neutral particles, the free energy density (i.e., the
parameters $\varepsilon=\gamma$, $\sigma$, and $\lambda$) and the corresponding
stability diagram {\em are not modified by an external electric field}. 

The dynamics of $C$ is thus driven by the free-energy density~(\ref{free}),
but, in addition, 
there is a {\em continuous creation of $C$ by the moving reaction front}, with 
a certain space and time dependent source density.
One is thus led to 
write down~\cite{antal99} the following phenomenological 
evolution equation for the reduced concentration field $m(x,t)$:
\begin{eqnarray}
\frac{\partial m}{\partial t}&=&-\lambda \Delta \frac{\delta {\cal F}[m]}{\delta m} + S_m(x,t)\nonumber\\
&=&-\lambda\Delta (\varepsilon m -\gamma m^3+\sigma \Delta m) + S_m(x,t)\,,
\label{CH}
\end{eqnarray}
where $\lambda$ is a kinetic coefficient and $S_m(x,t)$ is the source density.
Equation~(\ref{CH}) has to be solved with the homogeneous initial condition 
$m(x,t=0)=-(c_h+c_l)/(c_h-c_l)$ that corresponds to the absence 
of $C$ inside the system before the beginning of the reaction. 
Note that the above ``Cahn-Hilliard equation with a source" 
should also contain two noise terms. One of them is the thermal noise, 
while the other one  originates in the chemical reaction that creates the source term. 
However, as discussed in~\cite{cornell93}, 
the noise in  $A+B\rightarrow C$-type reaction
fronts can be neglected in dimensions $d \geq 2$, 
while neglecting the thermal noise term
means that an effective zero-temperature process is considered, 
and that the 
phase separation takes place {\em only} through a spinodal decomposition mechanism. This approximation 
is supported by the experimentally-known fact of the very long life of the formed patterns, 
which amounts to a very low `effective temperature' of the system. The 
theory could be refined 
by including the thermal noise, since then the nucleation and growth 
processes would be also captured. The role of noise has been investigated for
the fieldless case in~\cite{antal01}, and its effects in the presence of an
electric field will be the subject of a forthcoming paper. 
Here we shall remain within the deterministic framework corresponding to
Eq.~(\ref{CH}). 

Let us now concentrate on the source term $S_m(x,t)$ in Eq.~(\ref{CH}). 
As already mentioned, 
this term models the production of $C$ by the moving reaction front,
and it is {\em influenced by the presence of an external electric field}.
The effect of the electric field has been studied in detail in~\cite{bena05},
and we summarize below the main results
{\em for the range of parameters that are relevant 
for typical experimental situations}.
One has to realize first that the reagents $A$ and $B$ are electrolytes
which dissociate,
\begin{equation}
A \rightarrow A^++ A^-\,,\qquad B \rightarrow B^++B^-\,,
\label{dissociation}
\end{equation}  and the basic reaction process is
\begin{equation}
A^-+B^+\rightarrow C \,,
\label{reaction}
\end{equation}
while the `background' ions $A^+$ and $B^-$ are not reacting.
The modelling of the system in~\cite{bena05} was based on 
several simplifying assumptions:
(i) the one-dimensional character of the system
(i.e., all the relevant parameters are only $x$-dependent); 
(ii) the complete dissociation of the 
electrolytes $A$ and $B$ into their respective ions 
(that allows to eliminate the dynamics of the 
neutral $A$-s and $B$-s from our description); 
(iii) infinite reaction rate and irreversibility of the 
basic reaction $A^-+B^+\rightarrow C$. This is justified by the fact that 
the characteristic reaction time-scale
is much smaller than any time-scale connected with diffusion and 
precipitation (pattern formation), 
and leads to a point-like reaction zone; (iv) the electroneutrality  
approximation
(the local charge density is zero on space scales that are 
relevant to pattern formation), 
whose applicability for
the systems under study was discussed in detail in~\cite{unger00}; 
(v) we considered monovalent ions; 
(vi) finally, we assumed equal diffusion coefficients $D$ of the ions. 
As mentioned in~\cite{bena05}, 
any of these assumptions may be relaxed without 
generating major changes of the conclusions of our study. 
The reaction-diffusion equations for 
the concentration profiles of the ions can be solved numerically 
(with boundary conditions that 
correspond to the presence of the two reservoirs of ions 
-- of concentrations $a_0$ and $b_0$, respectively -- at the ends of the reaction cylinder).

A first result  refers to the motion of the point-like reaction zone; 
it is, to a good approximation, a diffusive motion, 
\begin{equation}
x_f(t)=\sqrt{2D_ft}\,,
\end{equation}
with a diffusion coefficient 
$D_f$ that is practically unaffected by the field intensity, i.e., 
it is given by its fieldless expression
\begin{equation}
\mbox{erf}\left(\sqrt{\frac{D_f}{2D}}\right)=\frac{(a_0/b_0)-1}{(a_0/b_0)+1}\,,
\label{dfront}
\end{equation}
where $D$ is the common diffusion coefficient of the ions. Note that for a
reverse field there is also a small drift component in the motion of the
front. However, for the range of reverse fields and observation times 
we are considering, this drift component can be neglected.

The effect of the field is more significant in the production of $C$. 
As known~\cite{antal98}, in the absence of the field the concentration 
of $C$-s left 
behind the front is a constant and its value $c_0$ is determined by the initial
concentrations of the ions  $a_0$ and
$b_0$, and by their diffusion coefficients. 
In the particular case of equal diffusion coefficients $D$ of the ions,
its value is given by:
\begin{equation}
c_0 \approx  a_0\,K\sqrt{2D/D_f}\,,
\label{c0}
\end{equation}
where $K\equiv (1+b_0/a_0)(2\sqrt{\pi})^{-1}\,\exp(-D_f/2D)$, 
and the diffusion
coefficient $D_f$ of the front is given by Eq.~(\ref{dfront}).
For an infinite reaction rate, the corresponding source
density for the production of $C$ (i.e., the density of production 
of $C$ per unit time)
is a $\delta$ peak localized at the instantaneous position of the 
front $x_f(t)$,
\begin{equation}
S_m(x,t)=\frac{a_0\, K \,\sqrt{D}}{(c_h-c_l)/2}\,\,
\frac{\delta(x-x_f(t))}{\sqrt{t}}
\label{c-source}
\end{equation}
whose amplitude decays in time as $\sim 1/\sqrt{t}$.

\begin{figure}[h!]
\begin{center}
\vspace{1.5cm}
\includegraphics[width=0.9\columnwidth]{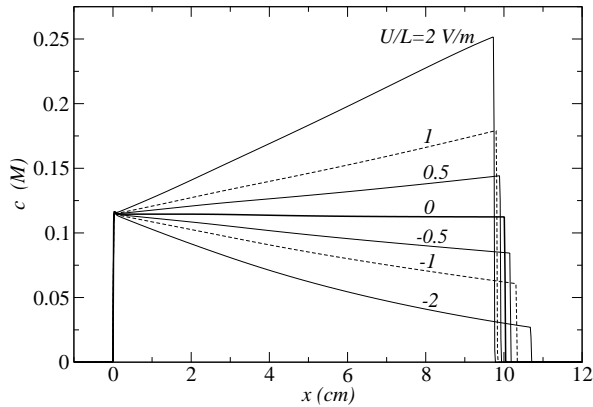}
\end{center}
\caption{The density of the reaction product $C$
left behind the reaction front for different values
of the electric field $U/L$ applied to the system.
The observation time is $t=10$ days.
The values of the other parameters are: $a_0=10$ M,
$b_0=0.1$ M, and $D=10^{-9}$ m$^2$/s, that lead to $c_0=0.1145$ M.}
\label{figure3}
\vspace{1.5cm}
\end{figure}

Consider now an external electric field applied to the system.
For both polarities of the field, 
and for relatively small values of the field intensity (the so-called {\em linear regime}
at which we shall limit our study in this paper), there is a linear variation of the concentration
of $C$-s with $x$, with a slope that is proportional to the applied tension, 
\begin{equation}
c(x)=c_0[1 + (\eta \,U/L) x]\,,
\label{cx}
\end{equation}
as illustrated by Fig.~\ref{figure3}. The parameter $\eta$ 
depends on the values of the other parameters 
(i.e., $a_0$, $b_0$, and $D$), and for the set of parameters
in Fig.~\ref{figure3}
one infers $\eta \approx 5$ V$^{-1}$. 
We should mention that the validity of this linear regime 
is wider for positive tensions
(e.g., it may go up to $U/L\approx 10$ V/m for the system considered in
Fig.~\ref{figure3})
and less extended for negative tensions 
(e.g., up to $U/L\approx -2$ V/m for the system in Fig.~\ref{figure3});
beyond these limits there is a relative error larger than $10$ \% in
approximating $c(x)$ by a linear profile with the initial slope.

The above result on the spatial dependence~(\ref{cx}) of the concentration of
the reaction product is
incorporated into the source term through the following modification of its
amplitude:
\begin{equation}
S_m(x,t)=\frac{a_0\, K \,\sqrt{D}}{(c_h-c_l)/2}\,\frac{[1 + (\eta \,U/L) x]\,\delta(x-x_f(t))}{\sqrt{t}}\,.
\label{sm}
\end{equation}

As discussed in the Introduction,
in the absence of an electric field ($U=0$)
this spinodal decomposition scenario reproduces, in a simple and coherent way, 
all the generic laws of Liesegang patterns. Moreover, 
it contains very few parameters, which can be inferred from experimental data~\cite{racz99}, 
and thus has a predictive power. We expect to recover these qualities in the 
presence of an applied external electric field, as well.

The solution to the  Cahn-Hilliard equation~(\ref{CH}) 
with the source~(\ref{sm}), 
i.e., the profile of the reduced concentration $m(x,t)$ is 
obtained numerically. As already mentioned above 
and discussed in detail in~\cite{bena05}, we decided to 
focus our analysis on the situations that are experimentally relevant and, 
in particular, the choice of parameters 
intends to mimic real experimental situations. Namely, 
we considered concentrations of the reagents $a_0$ and $b_0$ 
in the range $10^{-2}-10$ M, length $L$ of the system of some 
tenths centimetres, and tensions $U$ applied between system's 
edges such that we are in the linear regime of the 
production of $C$, i.e.,  $U/L$ 
varies between $-2$ and $+10$ V/m. The common diffusion 
coefficient of the ions was chosen as $D =10^{-9}$ m$^2$/s 
for all the calculations. The parameters $\lambda$, 
$\varepsilon=\gamma$, and 
$\sigma$ of the free-energy-driven part of the dynamics of $C$ can be 
inferred from the experimental data as explained in the Appendix. 
Finally, the pattern formation was followed for a period of the
order of ten days of experimental observation time. 
In the following section we present the results that were 
obtained for a polarity $U>0$ of the applied field (the 
forward field), that favours the reaction, i.e., that drives 
the $A^-$ and $B^+$ ions towards the reaction zone. Section~IV will 
be devoted to the $U<0$ (the reverse field) case study.

\section{Pattern characteristics for a forward applied field}

\begin{figure}[h!]
\begin{center}
\vspace{1.5cm}
\includegraphics[width=0.9\columnwidth]{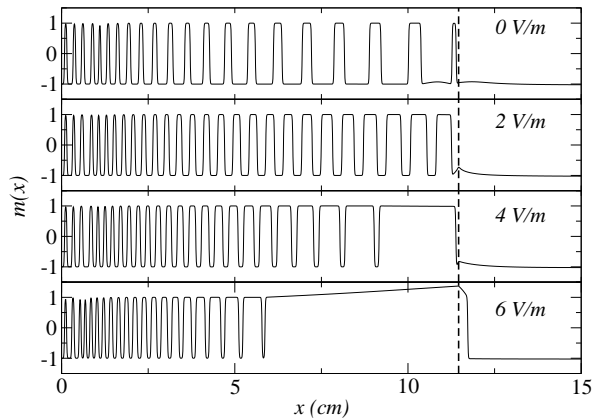}
\end{center}
\caption{The profile of the reduced concentration $m(x)$ for different values of the
forward applied field $U/L>0$. The snapshots are taken at $t=14$ days.
The dashed lines represent the position of the reaction front at this time.
The values of the other parameters are $c_0=0.1145\;M$, $c_l=0.0045\; M$,
$c_h=0.3645 \;M$, $D=10^{-9}$ m$^2$/s, $D_f=5.43 \cdot 10^{-9}$ m$^2$/s,
and $\eta=5$ V$^{-1}$.
One notices the decrease in the band spacing with increasing
tension, as well as the appearance of the plug -- earlier for
larger tensions.}
\label{figure4}
\vspace{1.5cm}
\end{figure}

A first generic feature of the patterns formed in the 
presence of the forward field ($U>0$ in our setup)
is that the number of bands 
is finite, i.e., band formation stops at a certain moment 
through the appearance of a continuous precipitation region 
(a `plug' of high-density precipitate). The higher 
the tension, the earlier this plug forms, see Fig.~\ref{figure4} for an 
illustration.

\begin{figure}[h!]
\begin{center}
\vspace{1.5cm}
\psfrag{l}{{\Large$\ell$}}
\includegraphics[width=0.9\columnwidth]{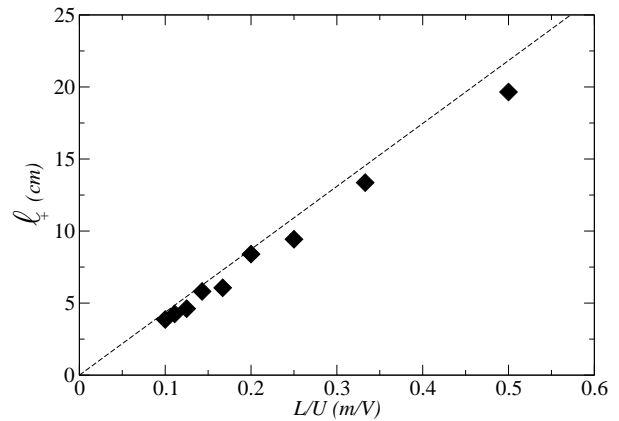}
\end{center}
\caption{The value of the distance $\ell_+$ of the plug-setting-in as a function of the
forward applied field $U/L>0$. The diamonds represent the numerically
estimated values, while the dashed line corresponds to the rough theoretical
estimate, Eq.(\ref{lplus}). The values of the parameters of the system are the
same as for Fig.~\ref{figure4}. The size of the symbols is representative for
the estimated error bars.}
\label{figure5}
\vspace{1.5cm}
\end{figure}

One can make a rough estimate of the distance $\ell_+$ of 
the onset of the plug through the following reasoning: the 
plug forms when the width of the $n$-th high-density band, 
$w_n$, becomes equal to the distance $(x_{n+1}-x_n)$ between 
the $n$-th and the $(n+1)$ high-density bands. If we estimate 
that the amount of $C$ produced by the front between $x_n$ 
and $x_{n+1}$ goes entirely in the $n$-th high-density band, i.e., 
\begin{equation}
c_0 [1 + (x_n + x_{n+1})/2](x_{n+1} -x_n)=c_h\,w_n = c_h\,(x_{n+1} - x_n)\,,
\end{equation} 
then we obtain for the distance $\ell_+ =(x_n + x_{n+1})/2 \approx x_n$: 
\begin{equation}
\ell_+=\frac{L}{\eta U}\;\frac{c_h-c_0}{c_0}\,.
\label{lplus}
\end{equation}
Figure~\ref{figure5} offers a comparison 
between the results of the numerical simulations 
and this theoretical estimation of $\ell_+$ for different values 
of the forward field $U/L>0$. Note that the experimental 
measurement of $\ell_+$ allows us to estimate the value of 
$c_h$ (provided that $c_0$ is known).

\begin{figure}[h!]
\begin{center}
\vspace{1.5cm}
\includegraphics[width=0.9\columnwidth]{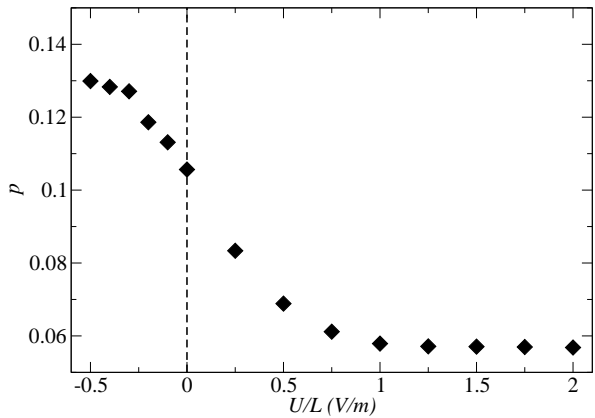}
\end{center}
\caption{The value of the effective spacing coefficient 
$p$ as a function of the
applied field  $U/L$. 
The values of the parameters of the system are the
same as for Fig.~\ref{figure4}. The size of the symbols is representative for
the estimated error bars.}
\label{figure6}
\vspace{1.5cm}
\end{figure}

A second generic feature of the pattern is that the distance 
between two successive bands diminishes as compared 
to the fieldless case, and this effect is increasing 
with increasing forward applied field $U/L$. This  can 
be easily understood through a simple qualitative argument. 
In the presence of the forward field, the reaction 
front leaves behind a larger quantity of $C$ than in the 
absence of the field. Thus, after the formation of a band, 
the re-establishment 
of the phase-separation instability conditions
takes place sooner in 
the presence of the field, resulting in a higher spatial density 
of bands in the system. 
With a good approximation, the positions of the bands 
$x_n$ still form a geometric series as in the fieldless case, and 
one can define an experimentally measurable `effective' 
spacing-law parameter $p$: 
\begin{equation}
x_n \sim (1 + p)^n \,.
\end{equation}
Thus, as illustrated in Fig.~\ref{figure6}, $p$ is a decreasing function of 
the forward applied field $U/L$. Due to 
the decrease in the number of bands with increasing field 
intensity our measures of $p$ were 
restricted to a rather narrow interval of field intensities 
around $U/L =0$. 

\begin{figure}[h!]
\begin{center}
\vspace{1.5cm}
\includegraphics[width=0.9\columnwidth]{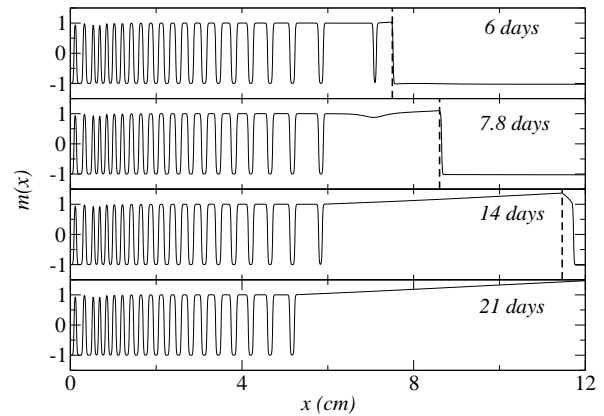}
\end{center}
\caption{The profile of the reduced concentration $m(x)$ at different times 
for a fixed forward applied field  $U/L=6\;V/m$. 
The dashed lines represent the position of the reaction front at 
the corresponding times. 
One notices the progressive `flow-back' of the $C$-s from 
plug zone, i.e., the backward disappearance of the pattern. 
The values of the parameters of the system are the
same as for Fig.~\ref{figure4}.}
\label{figure7}
\vspace{1.5cm}
\end{figure}

Another remarkable feature of the pattern is its `instability'. 
Indeed, as illustrated by Fig.~\ref{figure7}, due to the high 
densities of C in the plug region (which thus becomes unstable), 
there is a `flow-back' of C from the plug towards 
the regions where the patterns had formed previously. 
Therefore, this may cause a gradual disappearance of 
some of the already formed bands. This effect is not present in the fieldless
case.

\begin{figure}[h!]
\begin{center}
\vspace{1.5cm}
\includegraphics[width=0.9\columnwidth]{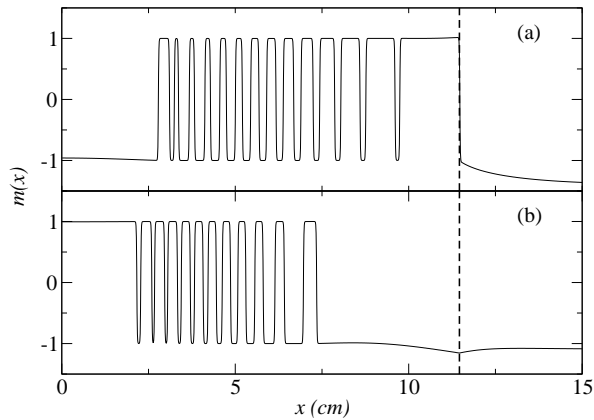}
\end{center}
\caption{The profile of the reduced concentration $m(x)$ at $t =14$ days for 
two sets of values of the parameters: (a) $c_0 =0.1145$ M, 
$D_f =5.43\cdot 10^{-9}$ m$^2$/s, $\eta=5$ V$^{-1}$,
$c_h =0.72$ M, and $c_l =0.120$ M 
(i.e., $c_0$ 
is below the spinodal decomposition domain), and an applied 
electric field $U/L = 10$ V/m. One notices the initial absence 
of pattern, then a limited region of existence of the pattern 
and, in the end, a plug region starting to form. (b) $c_0 = 
0.1145$ M, $D_f =5.43 \cdot 10^{-9}$ m$^2$/s, $\eta=5$ V$^{-1}$,
$c_h =0.1045$ M, 
and $c_l = 0.045$ M (i.e., $c_0$ is above the spinodal decomposition domain), 
and an applied electric field $U/L = -2$ V/m. One notices the 
initial plug region, then a limited region of existence of the 
pattern, and, in the end, the disappearance of the pattern. 
The dashed line represents the position of the front at the 
time of the snapshot.}
\label{figure8}
\vspace{1.5cm}
\end{figure}

One has to realize that the structure of the patterns 
depends essentially on the position of the
concentration  of $C$-s left behind by the reaction front 
with respect to
the stability domains  associated 
with the free energy density ${\cal F}$. 
This remark holds both in the absence and in the presence of an 
applied field. If, for example, $c_0$ is below the spinodal 
value, no pattern will form in the absence of the field. However, 
in the presence of a forward field, the concentration 
of $C$-s left behind the front is continuously increasing, 
and thus, at a certain moment, it will cross the spinodal 
curve, and therefore the pattern formation mechanism 
will be `turned-on', i.e., patterns will start to form after 
a certain time and then continue up to the eventual plug 
formation discussed above, as illustrated in  Fig.~\ref{figure8}a.

We can also comment here on the width law~\cite{Muller82,droz99,racz99}. 
As shown in~\cite{racz99}, in the fieldless case the derivation of this law, 
$w_n \sim x_n^{\alpha}$, with $\alpha=1$, is straightforward. 
Let us apply the same reasoning as in~\cite{racz99} in the presence of an electric field. 
One combines the facts that (i) the reaction front leaves behind a density $c(x)$
of $C$-s, Eq.~(\ref{cx}); (ii) the $C$-s segregate into low ($c_h$) and high ($c_h$)
density bands; (iii) the number of $C$-s is conserved in this segregation process.
The equation expressing the conservation of $C$-s can be written as
\begin{equation}
c_0\left[1+(\eta U/L) (x_n+x_{n+1})/2\right](x_{n+1}-x_n)=c_hw_n+ c_l(x_{n+1}-x_n-w_n)\,,
\end{equation}
which together with $x_{n+1}=(1+p)x_n$ leads to
\begin{equation}
w_n=\frac{p(c_0-c_l)}{c_h-c_l}\,x_n+\frac{c_0p(1+p/2)\eta U/L}{c_h-c_l}x_n^2\,.
\label{width}
\end{equation}
It is however difficult to make a clear-cut statement on this point. Indeed, for
field intensities for which the second term in~(\ref{width}) may start to play a
role, the total number of bands is so small that no reliable conclusion can be
drawn about the `systematic' behaviour of their width. In case one attempts
to fit $w_n\sim x_n^{\alpha}$, one will infer an effective $\alpha$ which is
between $1$ and $2$ and which increases with the field.

\section{Pattern characteristics for a reverse applied field}

\begin{figure}[h!]
\begin{center}
\vspace{1.5cm}
\includegraphics[width=0.9\columnwidth]{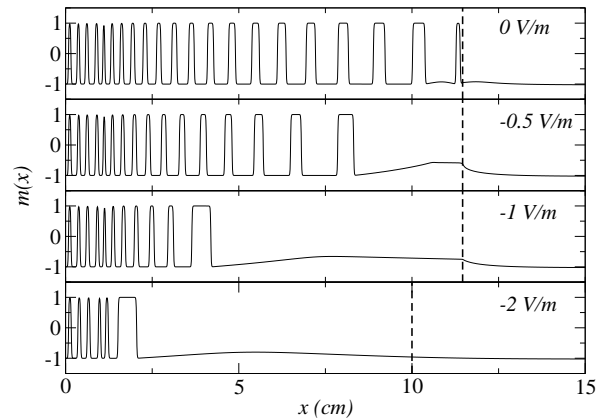}
\end{center}
\caption{The profile of the reduced concentration $m(x)$ for different values of the
reverse applied field $U/L<0$. The snapshots are taken at $t=14$ days.
The dashed lines represent the position of the reaction front at this time
(except for the last panel, where it indicates the stopping of the reaction,
$S_m=0$). One notices the increase in band spacing with decreasing
applied tension, as well as the rapid disappearance of the pattern.
The values of the parameters of the system are the
same as for Fig.~\ref{figure4}.}
\label{figure9}
\vspace{1.5cm}
\end{figure}

Let us consider now the case of the reverse polarity 
of the electric field (i.e., the case when it drives 
the reacting ions away from
the reaction zone, $U<0$ in our setup). 
Again, a generic characteristic of the pattern is the finite 
number of bands. This can be easily understood in 
connection with the continuous decrease of the concentration 
of $C$-s left behind the reaction front, up to a 
point when the phase-separation conditions are no 
longer fulfilled. As illustrated in Fig.~\ref{figure9}, there are less 
and less bands for larger and larger field intensities. But 
contrary to the forward field case, these bands are `stable' 
i.e., nothing analogous to the `flow-back' process in 
this case. Moreover, the last formed band collects progressively 
all the $C$-s left behind the front, i.e., its width increases 
slowly with time.

There exists thus a maximum distance $\ell_-$ of the 
spatial extension of the pattern, which can be estimated 
experimentally as the right edge of the right-most 
high-density band. A rough theoretical estimation of this 
length is given through the condition that the concentration 
of the $C$ product at this point equals the lower 
limit of the spinodal decomposition domain, i.e., 
$$c(\ell_-)=c_l+\left(1-\frac{1}{\sqrt{3}}\right)\,\frac{c_h-c_l}{2}\,,$$
from which
\begin{equation}
\ell_-=\frac{L}{\eta |U|}\;\frac{(c_0-c_l)-(1-1/\sqrt{3})(c_h-c_l)/2}{c_0}\,.
\label{lminus}
\end{equation}

\begin{figure}[h!]
\begin{center}
\vspace{1.5cm}
\psfrag{l(cm)}{{\Large$\ell_-$ (cm)}}
\includegraphics[width=0.9\columnwidth]{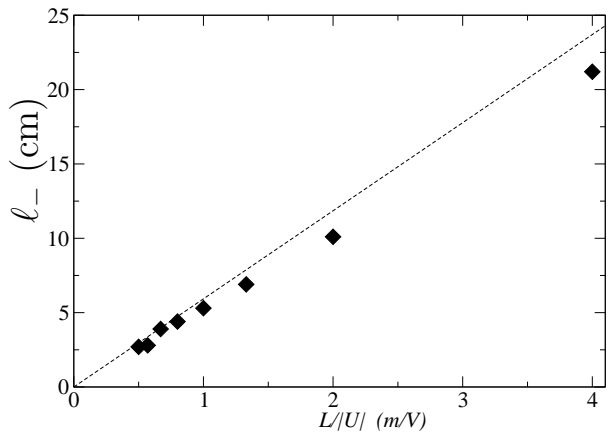}
\end{center}
\caption{The value of the distance $\ell_{-}$ of disappearance of the
pattern as a function of the
absolute value of the reverse applied field $|U|/L$. The diamonds represent the numerically
estimated values, while the dashed line corresponds to the rough theoretical
estimate, Eq.(\ref{lminus}).
The values of the parameters of the system are the
same as for Fig.~\ref{figure4}.  The size of the symbols is representative for
the estimated error bars.}
\label{figure10}
\vspace{1.5cm}
\end{figure}

A comparison of this result with the numerical simulations 
is given in Fig.~\ref{figure10} for different values of the 
field intensity $|U|/L$.

We can see that the expressions for the two lengths $\ell_+$ and $\ell_-$
(Eqs.~(\ref{lplus}) and ~(\ref{lminus})) provide
a way to evaluate $c_h$ and $c_l$ through experimental 
measurements of the length of the patterns.

As expected, in the reverse field case, there is an increase 
of the spacing between the bands (as compared to 
the fieldless case), and this is illustrated by an increasing 
of the effective spacing coefficient $p$ with increasing field 
intensity $|U|/L$, see Fig.~\ref{figure6} for a numerical estimation. 

Finally, it is worth mentioning again how the appearance 
of the pattern is sensitive to the position of $c_0$ with 
respect to the stability domains associated 
with the free energy density ${\cal F}$. If, e.g., $c_0$ is above this 
instability domain, there is no pattern formation in the 
fieldless case, but just a uniform plug of $C$ behind the 
front. However, as illustrated in Fig.~\ref{figure8}b, in the presence 
of a reverse field, due to the continuous decrease in 
the concentration of $C$, the instability domain may be 
reached from above, and a pattern starts to form after 
an initial plug region, and eventually, as discussed above, 
stops after some time, when the instability domain is left 
for too small concentrations of $C$.

\section{Comparison with experiments}

Experiments on Liesegang patterns in the presence of
an electric field have been going on quite a while since the 
pioneering works of Happel {\em et al.}~\cite{happel29}
and Kisch~\cite{kisch29}, 
see~\cite{ortoleva78,Feeney-Orto1983,sharbaugh89,das90,das91,das92,
sultan00,sultan02,ghoul03,lagzi02,lagzi03,shreif04}. 
As already mentioned in~\cite{bena05}, with the exception of~\cite{sharbaugh89,lagzi02,lagzi03}, 
these experiments were carried on for a single polarity of the electric field (i.e., either
a `forward', or a `reverse' field according to our terminology).~\footnote{There are some ambiguities in \cite{sharbaugh89} on the polarity of the applied field.} 
It is therefore  not surprising that each of them  `covers'
only a part of the situations that are predicted by our theory,
which thus has the merit to include all these elements in an unifying, coherent, and simple frame.
We shall present below a brief qualitative summary of these experimental results.

Concerning the motion of the front, it is experimentally found to be {\em diffusive} 
in~\cite{sharbaugh89},
\cite{das90} (with a slight decrease in the 
diffusion coefficient with increasing field intensity),
and~\cite{das91} (for a $2D$ geometry), 
in agreement with our 
theoretical predictions~\cite{bena05} for a 
{\em forward} field. Other experiments~\cite{sultan00,sultan02,ghoul03},
present a motion of the front with a small drift component 
(that increases with increasing field intensity~\cite{sharbaugh89}).
Occasionally, this drift component is given a theoretical justification
on the basis of a reaction-diffusion model for the system with a
superimposed constant electric field intensity, 
see~\cite{ghoul03,lagzi02,lagzi03,Feeney-Orto1983} --
that, as discussed in~\cite{bena05}, 
is a somewhat unrealistic assumption.
Here again our model is able to capture 
the appearance of this small drift component for the case of a
{\em reverse} electric field.

Several experimentally-studied modifications 
of the characteristics of the Liesegang patterns in 
the presence of an electric field
are quoted in the literature, and sometimes they seem contradictory:\\ 
(i) A first striking observation is the appearance of an
uniform precipitation for sufficiently high 
field-intensities~\cite{ortoleva78,sharbaugh89,das90,das91} 
and/or after a sufficiently long
time~\cite{sharbaugh89}. \\
(ii) Also, there is an `acceleration' 
of the band formation~\cite{sharbaugh89},
i.e., the time of appearance of the first band decreases 
with increasing field intensity~\cite{das91}.\\
(iii) A decrease of the spacing between successive bands 
(as compared to the fieldless case)
with increasing field intensity
is registered in~\cite{ortoleva78,das91}.\\
(iv) On the other hand, an increase of the spacing between 
successive bands is described
in~\cite{sultan00,sultan02,ghoul03},
and~\cite{shreif04} (for a $2D$ experimental setup). \\
(v) As reported in~\cite{sultan00,sultan02,shreif04},
at a given time there are less bands formed for higher field intensities. \\
(vi) The paper~\cite{das91} indicates a reduction of the initial 
``diffuse portion" with increasing field
intensity, and~\cite{shreif04} also gives an example of the 
reduction of the initial ``fuzzy zone".\\

One is now in position to recognize that all these features are 
recovered in a simple way within our theory.
More precisely, (i)-(iii) and the first result in (vi) 
are obtained in our model in the case of a forward
electric field, see Section~III; while (iv), (v), and 
the second result in (vi) can be obtained 
theoretically for a reverse applied field, 
see Section~IV.\\
(vii) Finally, a special remark on the results of~\cite{lagzi02,lagzi03}, which sometimes match our results, 
sometimes are just the opposite
of those obtained from our theory (as far as the polarity of the electric field is concerned).
In particular, in~\cite{lagzi02} 
it was found that (a) the motion of the reaction front is diffusive for a reverse field and has a small drift component for a forward field (contrary to our theory); (b) the average spacing coefficient $p$ decreases with increasing field (just like in our theory). In~\cite{lagzi03} an attempt to 
fit the width law~(see, e.g.,~\cite{racz99}) 
in the presence of an electric field,
$w_n\sim x_n^{\alpha}$, leads to an exponent 
$\alpha$ that is decreasing monotonically with the field  --
in opposition with the result of our discussion following Eq.~(\ref{width}). 
However it was argued~\cite{lagzi04} 
that in his case the properties of the intermediate compounds are
responsible for this `anomalous' behaviour, and such effects are outside 
the range of our theoretical model.

Most of the above examples show agreement with our theory at a qualitative level.
It should be noted however that our model offers {\em quantitative}
estimates of the changes as compared to the fieldless case, 
and thus is well-suited for direct comparison with the results of
appropriately-designed experiments.

\section{Conclusions and perspectives}

In this paper we examined the influence of an applied 
electric field on the formation and characteristics 
of Liesegang patterns that appear in the wake of an 
$A^- + B^+ \rightarrow C$ reaction front. The appearance of the 
pattern is related to the phase separation of the precipitate 
$C$ into a high and a low density phase. At a 
macroscopic, phenomenological level, the dynamics of $C$ 
particles can be modelled using a Cahn-Hilliard equation 
supplemented with a space and time-dependent source 
term that describes the production of $C$-s by the reaction 
front. It was found that for both polarities of the applied 
field the pattern has a finite spatial extension. Moreover, 
measuring this spatial extension for various values of the 
applied tensions allows to infer the values of the concentrations 
of $C$ in the high and low density phases. The 
distance between the bands in the pattern is influenced 
by the intensity of the applied field, namely it decreases 
with increasing $U/L$ for a forward field $U > 0$, and increases 
with increasing $|U|/L$ for a reverse field $U < 0$. 

One has to underline that the proposed model contains 
a minimal number of parameters, and all of them can 
be inferred from experimental data. Thus the model, 
besides reproducing well the generic experimental laws 
of Liesegang patterns, has a clear predictive power that 
can be used to control experimental situations. 
The data we are presenting are the result of numerical 
simulations, and are based on a set of simplifying 
assumptions. It is expected that most of these assumptions
do not affect our main conclusions. However, the details of the reaction process
leading to the final precipitate $C$ may be more complex  (e.g., 
implying several
steps, intermediate compounds with different 
electric charges and diffusivities,
etc.), and the role of the electric field may also have unexpected features.
The simplest extension of our work would be to consider a 
reaction scheme with one bivalent and two monovalent
ions, a case that occurs in several experiments. Work on this type of systems is in progress.  

\appendix
\section{Setting the parameters}

Typical experimental situations correspond to concentrations 
$a_0$ and $b_0$ of the reagents of the order of $10^{-2}-10$ ~M, 
with a ratio $a_0/b_0 \sim 10 -100$, and thus it is suitable 
to choose the unit of concentration as $n_0 =1$ M. 
The diffusion coefficients of the reagents are of the order 
$\sim 10^{-9}$~m$^2$/s, so that the length $\ell$ 
and time $\tau$ scales should be chosen 
such that $\ell^2/\tau$ is of the same order of 
magnitude, 
$$
\frac{\ell^2}{\tau} \sim D\,.
$$ 
Moreover, the experimental patterns have a total 
length of about $20$ cm, and the time to produce such a 
pattern is of some tenth days -- and this offers the order 
of magnitude of the diffusion coefficient of the reaction 
front, which is typically of the same order or an order 
of magnitude larger than the diffusion coefficients of the 
reagents (depending on the ratio $a_0/b_0$ of the concentration 
of the reagents). 
The typical widths of the precipitation bands are of 
a few mm at the beginning, and approach $\sim  1$ cm at 
the end, and so are the distances between two successive 
bands. From the visual observation of the beginning of 
the band formation it takes some tenth minutes for the 
band to be clearly seen, and some hundred minutes for its 
complete formation. Since the Cahn-Hilliard model has 
intrinsic length scale $\sqrt{\sigma/\varepsilon}$
and time scale $\sigma/(\lambda \varepsilon^2)$, 
these have to be comparable to the typical length and 
time scales of the appearance of a single band. 
We are thus led to the following orders of magnitude 
for the length and time scales: 
$$\ell=\sqrt{\sigma/\varepsilon}\sim 10^{-4}\,\mbox{m}\,,
\quad \tau =\sigma/(\lambda \varepsilon^2) \sim 10\,\mbox{s}\,. 
$$ 
There remains, however, the question of determining 
the concentrations of the two phases of the $C$ precipitate, 
$c_h$ and $c_l$.  As explained in the main text, this can be done 
through measurements of the total spatial extent of the 
pattern in the presence of a forward, respectively reverse 
electric field.

Here is the set of parameters that we 
used in most of the simulations presented above. For the 
concentrations of the reagents we take $a_0 =10$ M, $b_0 =0.1$ M, 
and the common diffusion coefficient of the electrolyte 
ions is $D =10^{-9}$ m$^2$/s. This leads, according to Eqs.~(\ref{dfront}) 
and (\ref{c0}), to a diffusion coefficient of the front 
$D_f = 5.43 \cdot 10^{-9}$ m$^2$/s, and a concentration of the fieldless  
reaction product $c_0 =0.1145$ M. For the above parameters the coefficient 
$\eta$ in the expression (\ref{cx}) of $c(x)$ is $\eta =5$ V$^{-1}$ . The 
parameters of the Cahn-Hilliard model are chosen such 
that $\ell=\sqrt{\sigma/\varepsilon}= 10^{-4}$~m, and 
$\tau =\sigma/(\lambda \varepsilon^2)=40$~s. 
Regarding the concentrations of the low and high density 
phases of $C$, we usually set them such that $c_0$ given above 
is in the instability domain of spinodal decomposition, 
$c_h =0.3645$~M, and $c_l =0.0045$~M.

\acknowledgments

We thank F. Coppex and I. Lagzi for very useful discussions.
This research has been partly supported by the 
Swiss National Science Foundation and
by the Hungarian Academy
of Sciences (Grants No.\ OTKA T043734 and TS 044839).

\end{document}